\documentclass[aps,twocolumn,nofootinbib]{revtex4} 
\usepackage[utf8]{inputenc}
\usepackage{graphicx}
\usepackage{braket}
\usepackage{dcolumn} 
\usepackage{bm} 
\usepackage{amsmath}
\usepackage{amssymb}
\usepackage{float}
\usepackage{physics}
\usepackage{soul}
\usepackage{hyphenat}
\usepackage{cancel}
\usepackage{thmtools}
\usepackage{bbding}
\usepackage[pdftex]{hyperref}
\hypersetup{colorlinks,linkcolor={blue},citecolor={blue},urlcolor={blue}}
\usepackage{subfigure}

\begin{document}
	\title{Wavefunction extreme values statistics in Anderson localization}
	\author{P. R. N. Falc\~ao}
	\email{pedro.falcao@fis.ufal.br}
	\affiliation{
		Instituto de F\'{i}sica, Universidade Federal de Alagoas, 57072-900 Macei\'{o}, AL, Brazil}
	\affiliation{
		Institute of Theoretical Physics, Jagiellonian University in Krakow, ul. Lojasiewicza 11, 30-348 Kraków, Poland}
	\author{M. L. Lyra}
	\email{marcelo@fis.ufal.br}
	\affiliation{
		Instituto de F\'{i}sica, Universidade Federal de Alagoas, 57072-900 Macei\'{o}, AL, Brazil}
	
	\begin{abstract}	
		We consider a disordered one-dimensional tight-binding model with power-law decaying hopping amplitudes to disclose wavefunction maximum distributions related to the Anderson localization phenomenon. Deeply in the regime of extended states, the wavefunction intensities follow the Porter-Thomas distribution  while their maxima assume the Gumbel distribution. At the critical point, distinct scaling laws govern the regimes of small and large wavefunction intensities with a multifractal singularity spectrum.  The distribution of maxima deviates from the usual Gumbel form and some characteristic finite-size scaling exponents are reported. Well within the localization regime, the wavefunction intensity distribution is shown to develop a sequence of pre-power-law, power-law, exponential and anomalous localized regimes. Their values are strongly correlated, which significantly affects the emerging extreme values distribution. 
		
	\end{abstract}
	\maketitle
	
	\section{Introduction}
	The Anderson localization transition is a key physical phenomenon related to a drastic change in the spatial distribution of Hamiltonian eigenfunctions promoted by disorder \cite{anderson1,anderson2,anderson3}. 
	In general lines, the eigenfunctions become exponentially localized in the regime of strong disorder while remaining spatially extended for weak disorder. Anderson transition is a quite general phenomenon affecting the transport of electronic, acoustic, magnetic and optical waves \cite{acoustic1,acoustic2,magnetic1, magnetic2,optical1}.

	The critical behavior of the Anderson transition has been extensively studied over the past decades and shown to strongly depend on the system's dimensionality, symmetries and the short or long-range character of the underlying couplings \cite{anderson3}. In particular, all eigenstates become exponentially localized in one-dimensional systems with short-range couplings for any finite amount of uncorrelated disorder. In contrast, one-dimensional systems with power-law decaying couplings support an Anderson transition and have been frequently used as a prototype model to investigate its universality classes and critical behavior \cite{long1,long2,long3,long4,long5,long6,long7,long8,long9,long10,long11,long12,long13,long14,long15,long16}. 
	
	Field theoretical renormalization group and random matrix (RM) theories yield several relevant aspects of the Anderson localization transition. In particular, random matrix theory unveiled universal characteristics of the eigenvalues statistics in the localized and extended phases \cite{rm}.
	Exponentially localized states are uncorrelated and randomly distributed along the energy band, resulting in a Poissonian probability distribution function (PDF) of the level spacements. On the other hand, the level repulsion typical of spatially extended states leads to a new spacement PDF that depends of the nature of the random matrix ensemble (Gaussian orthogonal, unitary or symplectic). RM theory has also been used to explore the statistical properties of the extreme eigenvalues \cite{extremeanderson}.
	
	Extreme events play a fundamental role in the study of disordered physical systems \cite{extreme1,extreme2}.
	These are events with magnitude much larger than the average value of a physical stochastic process. 
In condensed matter physics, they have a key impact on transport phenomena where the largest energy barriers and maximally localized modes hinder the transmission of physical excitations \cite{extremecondensed1,extremecondensed2}. 
The extreme values statistics of uncorrelated and identically distributed (IID) random variables is well understood \cite{extreme}.
Unbounded random variables having a PDF with a faster than power-law tail have a Gumbel distribution of the extreme values in long sub-sequences. A Fr\'echet extreme values PDF is in order for random variables with power-law decaying PDF tails. In the case of bounded stochastic variables, the extreme values are distributed according to an asymptotic Weilbull PDF. For sequences of strongly correlated random variables, very little is known regarding the extreme values statistics (for a recent review see \cite{extreme2}). In the context of Anderson localization, although RM theory discloses the PDF of the extreme values for the correlated eigenvalues of Gaussian ensembles \cite{extremeanderson}, studies of the extreme values statistics of the own eigenfunction intensities are missing.  Although the structureless nature of Gaussian extended eigenfunctions with an exponentially decaying Porter-Thomas intensity distribution\cite{porter} allows us to anticipate that the maximum distribution of extended states shall fall in the Gumbel class, multifractal correlations present in critical states as well as the structured aspect of exponentially localized states point to new distributions of the wavefunction intensity maxima that are still unexplored. 

In this work, we address the above relevant open question aiming to unveil how the extended or localized nature of single-particle eigenstates in a disordered system is reflected in the extreme values statistics of their eigenfunctions. We will consider the prototype one-dimensional Anderson model with random power-law decaying hopping amplitudes that exhibits a well-known localization-delocalization transition. We will focus in unveiling the scaling behavior of the average maximum value of the eigenfunctions as well as its probability distribution in the extended, critical and localized regimes. Further, we will unfold the role played by intrinsic eigenfunction correlations in the extreme values distributions.      

\section{Model and numerical methods}
We consider the following tight-binding Hermitian Hamiltonian model with power-law decaying hopping amplitudes on a linear chain with $N$ sites and periodic boundary conditions: 
\begin{equation}
	H = \sum_i \epsilon_i\ket{i}\bra{i} + \sum_{i > j} t_{ij}(\ket{i}\bra{j}+\ket{j}\bra{i}),
	\label{Eq.1}
\end{equation}
with $\epsilon_i$ representing the on-site potentials and hopping amplitudes given by $t_{ij} = W_{ij}/r_{ij}^{\sigma}$. $r_{ij}$ is the  distance between the chain sites, restricted to the interval $1\le r_{ij} \le N/2$  ($r_{ij}= i-j$ for $i-j<N/2$ and $r_{ij} = N-(i-j)$ for $i-j>N/2$). $\sigma$ is a characteristic power-law exponent that controls the effective range of the hopping amplitudes. We consider a large ensemble of the above Hamiltonian with $\epsilon_i$ and $W_{ij}$ being random real numbers distributed uniformly in the interval $[-1,1]$.  

The above Hamiltonian belongs to the class of power-law random band models. A perturbation analysis based on a field-theoretical model of interacting supermatrices, supported by exact diagonalization results, has settled that a well-defined Anderson transition occurs as a function of the control exponent $\sigma$ \cite{long1,long2,long3}. All states are extended for $\sigma <1 $. In this regime, the statistical properties are those of the Gaussian orthogonal ensemble of RM for $\sigma < 1/2$, with stronger fluctuations developing for $1/2< \sigma < 1 $. All states are critical at 
$\sigma =1$, exhibiting a multifractal character \cite{multifractal}. For $\sigma > 1$ all states are localized with integrable power-law tails $|\phi(r)|^2\propto r^{-2\sigma}$. A superdiffusive short-time spreading of wavepackets sets up for $1 < \sigma < 3/2$. 

Here, we will perform a statistical analysis of the extreme values of the above Hamiltonian eigenfunctions. Exact numerical diagonalization will be employed to compute the full spectrum of eigenvalues and the corresponding eigenfunctions of a large ensemble of disorder configurations. After a preliminary analysis of the density of states and level spacing distribution of the different regimes, we will focus on the eigenfunction statistics. One can associate a sequence of random variables to each eigenfunction $\{\phi_1^2, \phi_2^2, ..., \phi_N^2\}$, from which the distribution of eigenfunction intensities $P(\phi^2)$ can be extracted. We further identify the maximum $\phi_\mathrm{m}^2$ of each eigenfunction and its distribution $F(\phi_\mathrm{m}^2)$ over a spectral range around the band center. 

In the regime where the eigenfunction intensities are uncorrelated, the above two distribution functions are closely related. The cumulative distribution of maxima of IID eigenfunction intensities can be expressed as
\begin{equation}
	Q_N(\phi_\mathrm{m}^2) = \left[1-\int_{\phi_\mathrm{m}^2}^1 P(\phi^2)d\phi^2\right]^N ,
	\label{Eq.2}
\end{equation}
from which the distribution of maxima can be derived as $F(\phi_\mathrm{m}^2) = dQ_N/d\phi_\mathrm{m}^2$. Three limiting distributions of maxima can be anticipated for IID variables \cite{extreme}. In particular, when the tail of the parent distribution $P(\phi^2)$ decays faster than a power-law and the upper cutoff can be ignored, the distribution of maxima converges to the Gumbel form
\begin{equation}
	F (\phi_\mathrm{m}^2)\propto \exp[-a\phi_\mathrm{m}^2-b\exp(-a\phi_\mathrm{m}^2)],
	\label{Eq.3}
\end{equation}
where $a$ and $b$ are distribution parameters related to the average value and dispersion. Parent distributions with power-law tails lead to a distinct Fr\'echet distribution of maxima. On the other hand, when the parent distribution vanishes at and above a cutoff value, an asymptotic Weibull distribution of maxima of  IID sequences of random variables is predicted. In what follows, we directly compute both the parent distribution $P(\phi^2)$ and the maxima distribution $F(\phi_\mathrm{m}^2)$ of extended, critical and strongly localized states, with particular emphasis on possible deviations from the prediction based on the assumption of IID variables. 

\begin{figure}[h!]
	\centering
	\includegraphics[width = 0.49 \textwidth]{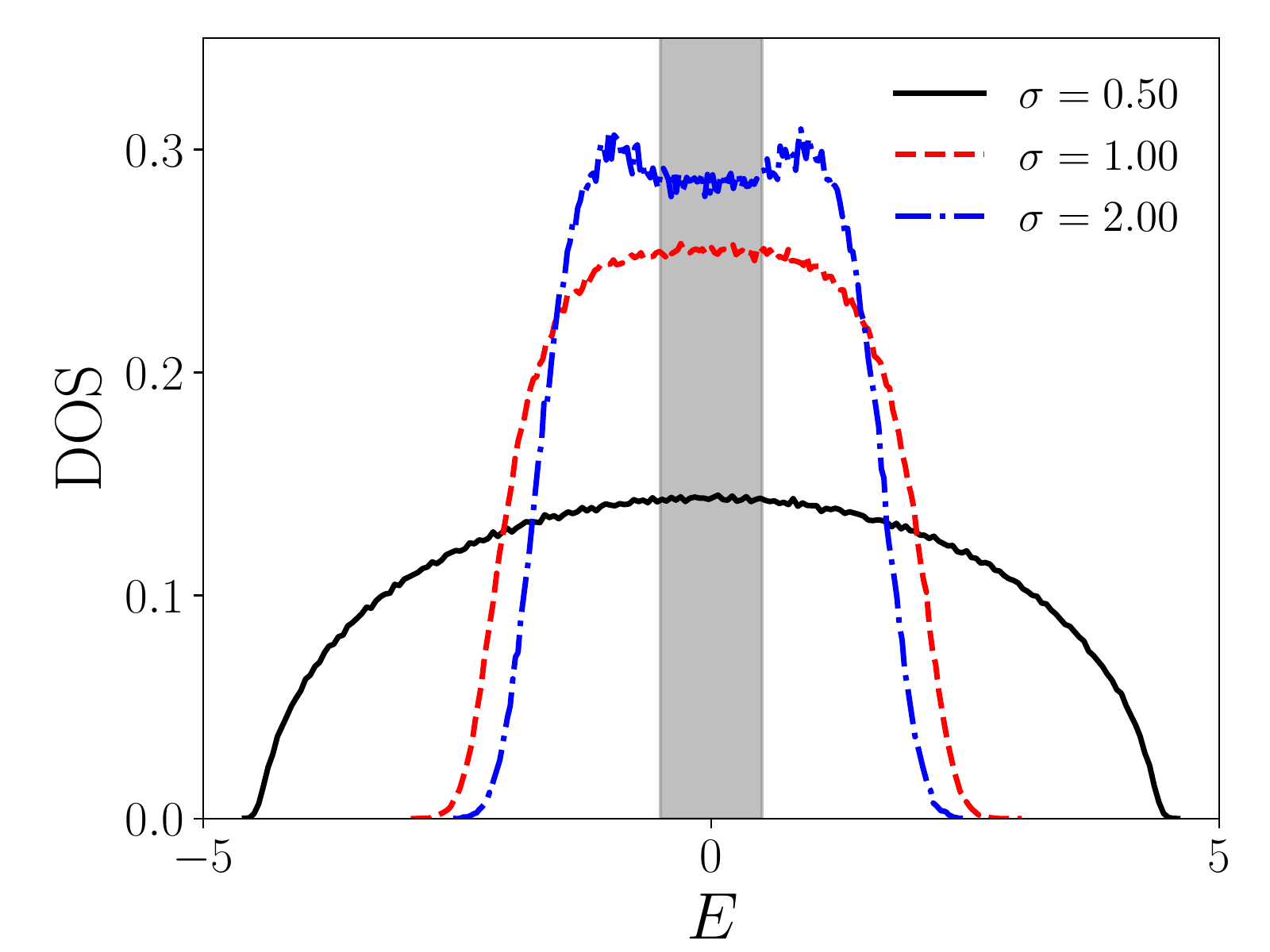}
	\caption{Density of states for three values of the hopping decay exponent $\sigma$ representing the extended ($\sigma =0.5$), critical ($\sigma = 1$) and localized ($\sigma=2$) regimes. The smooth profile of extended states evolves to a rough profile typical of localized states.  The shaded region represents the band central region $|E|\leq 0.5$, where the DOS has a flat profile. Data are from $10^3$ disorder configurations on closed chains with $N=10^3$ sites.}
	\label{Fig.1}		
\end{figure}
\section{Results}
We start displaying the density of states (DOS) for three representative values of the power-law exponent $\sigma$. Data shown in Fig.\ref{Fig.1} were obtained from the exact diagonalization of the matrix Hamiltonian on chains with $N=10^3$ sites and over an ensemble of $10^3$ disorder configurations. The DOS changes from the smooth profile at small $\sigma$ values, typical of effectively high dimensional systems with extended states, to a rough profile signaling the localized nature of states. The later is reminiscent from the DOS of the one-dimensional tight-binding model with short-range couplings on which the band edges singularities were rounded off by disorder. We highlight the central region of the spectrum in the energy range $E=[-0.5, 0.5]$ on which the DOS has a nearly flat profile, such that the average level spacing is roughly constant. 

\begin{figure}[t!]
	\centering
	\includegraphics[width = 0.49 \textwidth]{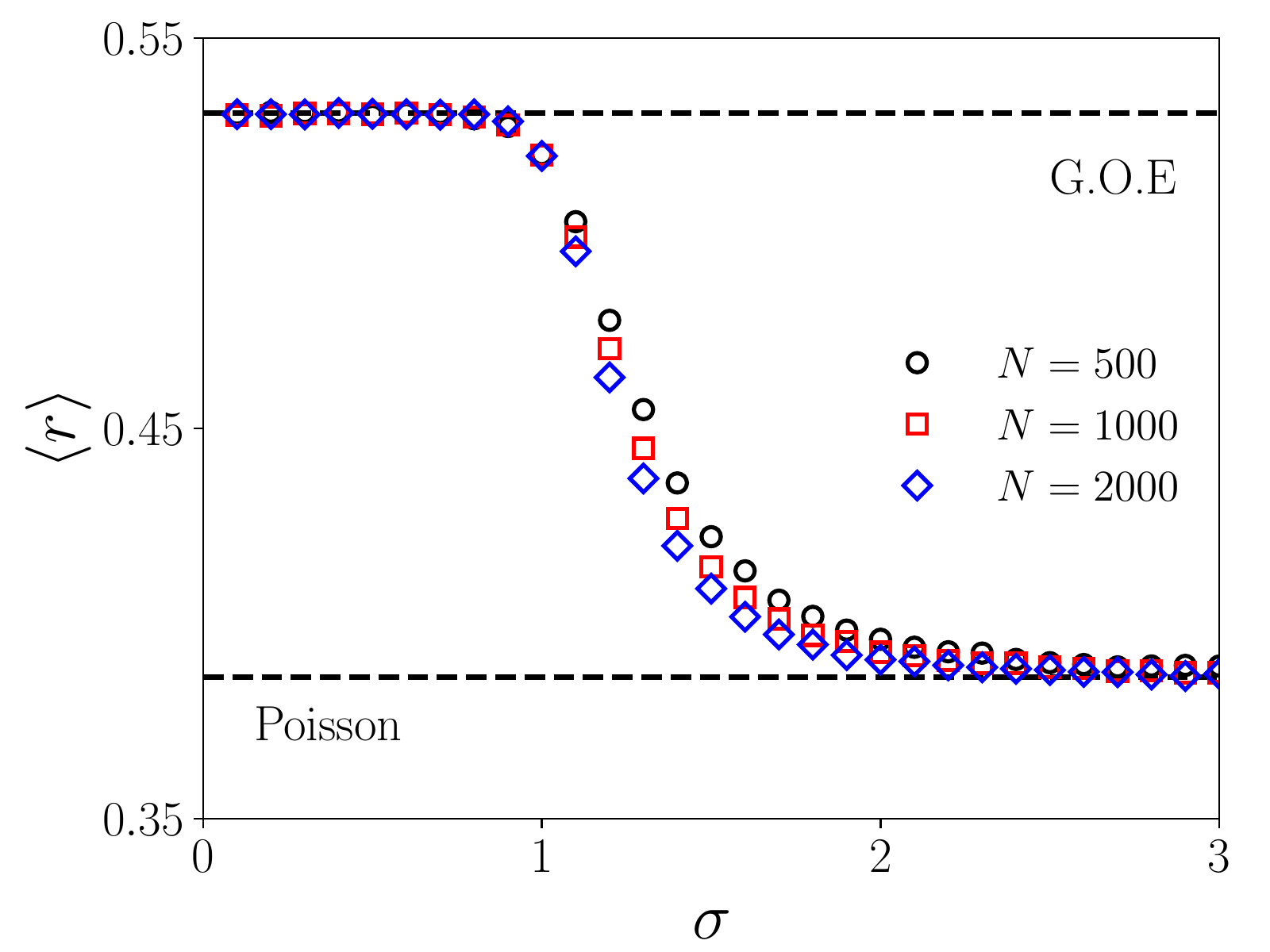}
	\caption{Average gap ratio $\langle r \rangle$ versus the 
		hopping exponent $\sigma$. Average was performed over all eigenstates in chains with distinct chain sizes, considering $10^3$ configurations of disorder.  Below the critical point ($\sigma = 1$), the gap ratio assumes the GOE value, with $\langle r \rangle \approx 0.5307$. In contrast, it slowly converges to the Poisson prediction for strongly localized states $\langle r \rangle = 2\ln2 - 1 \approx 0.3863$.  }
	\label{Fig.2}		
\end{figure}

To clearly evidence the transition from extended to localized states, we computed the average energy gap ratio \cite{gapratio1,gapratio2} defined as $\langle r\rangle =\langle \text{min}(s_n,s_{n-1})/\text{max}(s_n,s_{n-1})\rangle$, where $s_n = E_{n+1} - E_{n}$, as a function of $\sigma$, shown in Figure \ref{Fig.2}. For $\sigma< 1$, it assumes the value expected for extended states of a Gaussian orthogonal RM ensemble $\langle r\rangle\simeq 0.5307$. It slowly converges to the Poisson limit $\langle r\rangle =2\ln{2} - 1$ at large $\sigma$.

\begin{figure}[t!]
	\centering
	\includegraphics[width = 0.49 \textwidth]{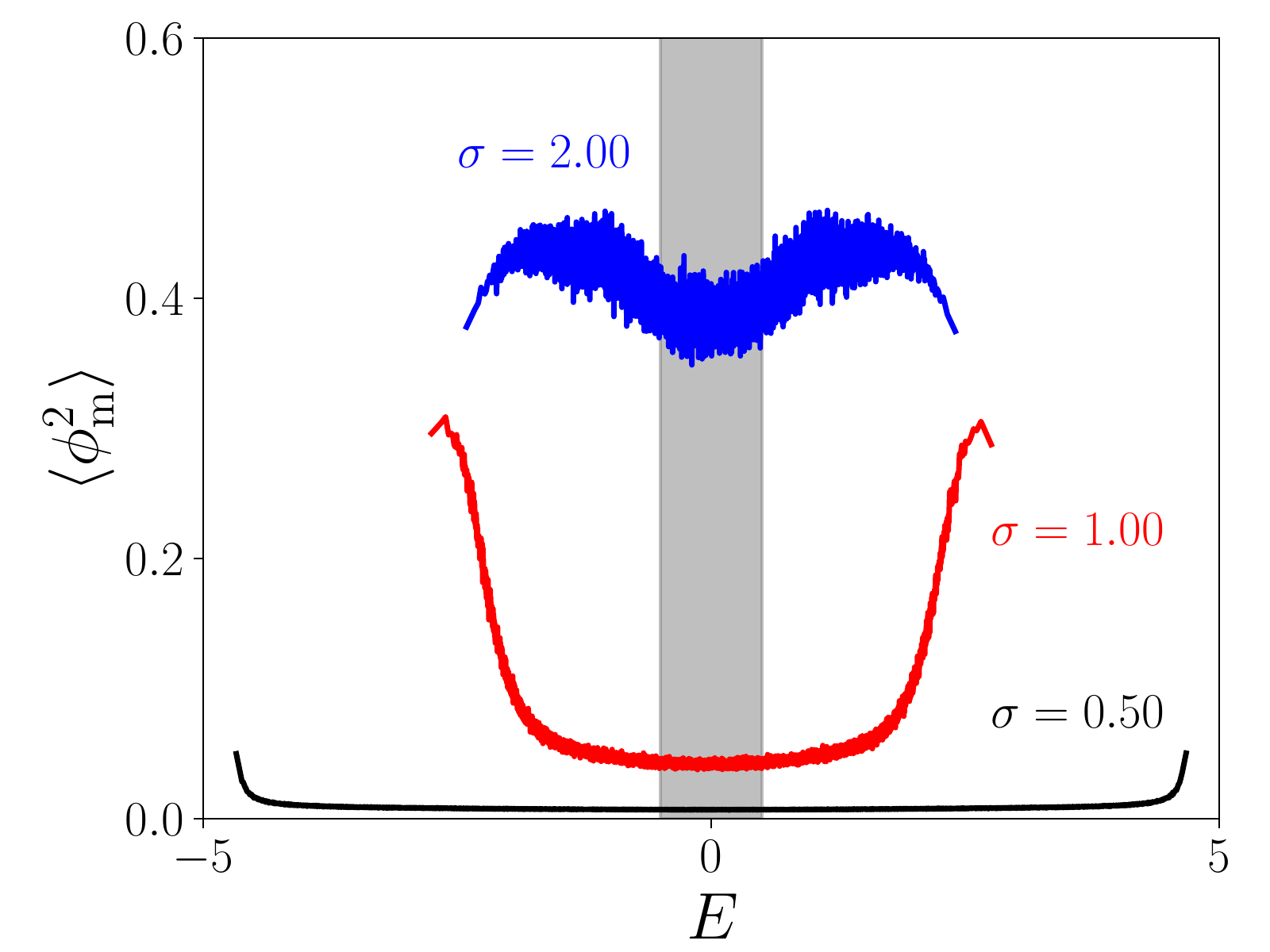}
	\caption{Spectrum of eigenfunction maximum intensity for the same values of $\sigma$ in Fig. \ref{Fig.1}. Shown values are averages over small energy windows. Notice the transition from extended (small maxima) to localized (large maxima) states. Near the band center the maxima are uniformly distributed. Data are from $10^3$ disorder configurations on chains with $N=10^3$ sites.}
	\label{Fig.3}		
\end{figure}

\begin{figure}[t!]
	\centering
	\includegraphics[width = 0.49 \textwidth]{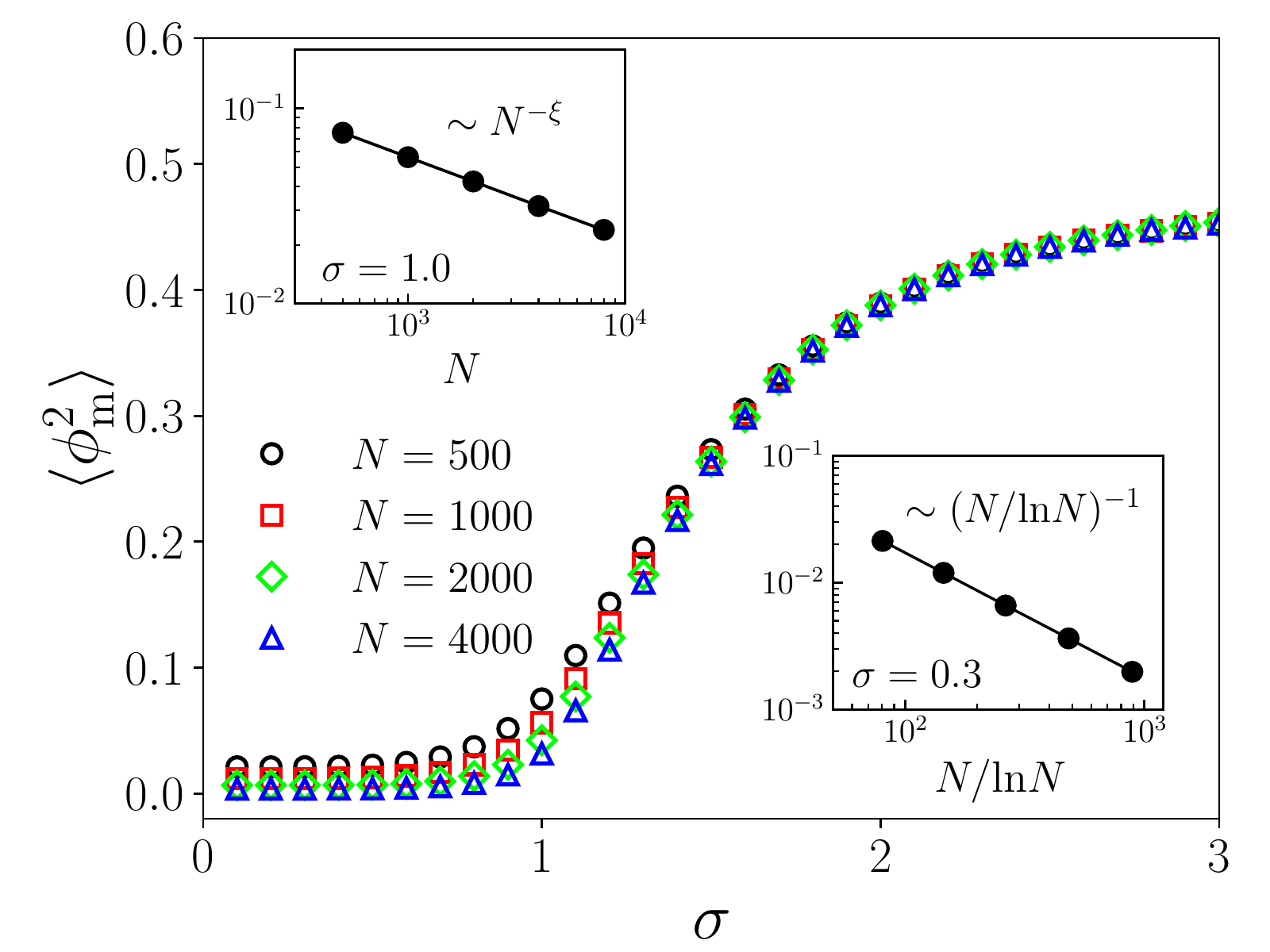}
	\caption{ Average eigenfunction maximum intensity as a function of the hopping exponent $\sigma$ for several lattice sizes. It behaves as an order-parameter measure, vanishing in the extended phase and becoming finite in the localized phase. Upper inset: Size dependence of $\langle \phi_\mathrm{m}^2 \rangle$ at the critical point. It displays a power-law decay $\langle \phi_\mathrm{m}^2\rangle \propto N^{-\xi}$, with $\xi = 0.413(2)$. Lower inset:  Size dependence of $\langle \phi_\mathrm{m}^2 \rangle$ deep in the extended phase ($\sigma = 0.3$). It decays as $\langle \phi_\mathrm{m}^2\rangle \propto \ln{N}/N$.}
	\label{Fig.4}		
\end{figure}

To study the extreme values statistics, we captured the maximum intensity of all eigenfunctions. Their average values on small energy windows are reported in Figure \ref{Fig.3} within the entire spectral range. In the phase of extended states, the average maximum decreases with the system size , showing reduced fluctuations. In contrast, they become size independent with large fluctuations for strongly localized states. In the highlighted central spectral window, the maxima are uniformly distributed. The average value of the intensity maxima can be used as an order-parameter measure of the Anderson transition. Figure \ref{Fig.4} brings its dependence on the control exponent $\sigma$. Deeply in the extended phase, the wavefunction maximum average vanishes in the thermodynamic limit as $1/N$ with a logarithmic correction ($\langle \phi_\mathrm{m}^2\rangle \propto \ln N/N$), as shown in the inset of figure \ref{Fig.4}. In opposite, it becomes finite in the localized phase. At the critical point, it assumes a power-law finite-size scaling $\langle \phi_\mathrm{m}^2\rangle \propto N^{-\xi}$ with $\xi \simeq 0.413(2)$.

\begin{figure}[p]
	\centering
	\includegraphics[width = 0.49 \textwidth]{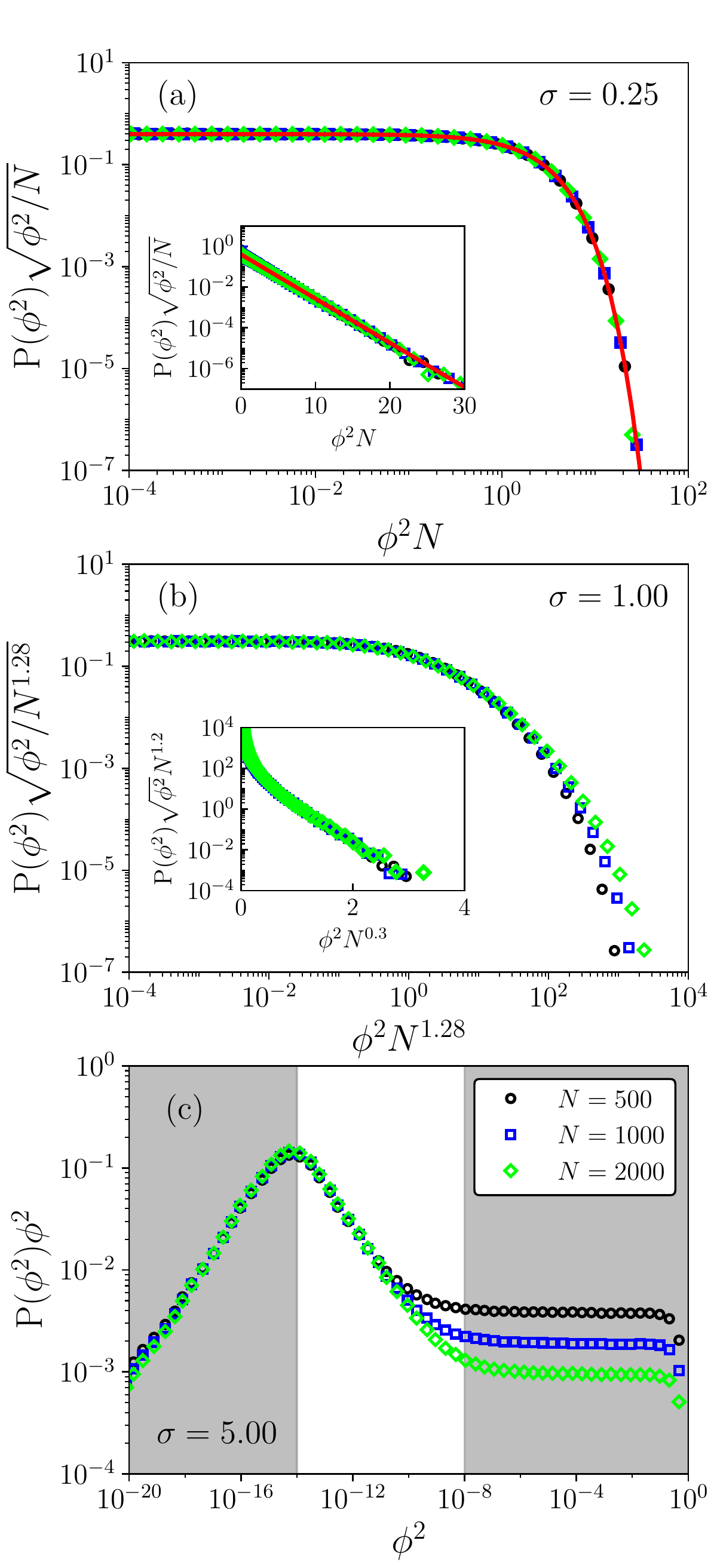}\\
	\caption{Scaled PDFs of the wavefunction intensities  for three representative values of $\sigma$. (a) $\sigma=0.25$ well within the extended phase. The wavefunction intensities follow the Porter-Thomas distribution (Eq.~4). Inset shows the same data in log-linear scale. (b) $\sigma=1$ (at the Anderson transition). For weak intensities, the critical distribution behaves as $N^{\beta}/\sqrt{\phi^2}$ with $\beta=0.64$. Inset shows the tail of the scaled critical PDF.  Notice, the slow convergence to the exponential form. Scaling exponents are distinct in the small and large intensities regimes. (c) $\sigma=5$ (well within the localized phase). Here we plot $\phi^2 P(\phi^2)$ to clearly evidence the four distinct regimes: extreme low intensities with $P(\phi^2)\propto 1/\sqrt{\phi^2}$ (first shaded region), the power-law regime with the $P(\phi^2)$ decaying faster than $1/\phi^2$ (white region), the exponential regime with $P(\phi^2)\propto (N\phi^2)^{-1}$ (plateau in the second dashed region) and the anomalously localized regime for the largest intensities (deep in the second shaded region).}
	\label{Fig.5}		
\end{figure}

Before addressing the distribution of maxima, we raised the parent distribution of the own eigenfunction intensities $P(\phi^2)$ built from all eigenfunctions with $-1/2<E<1/2$ \cite{long3,parent1,parent2}.  We start building $P(\phi^2)$ well within the extended phase where the GOE statistics are expected to hold. The resulting PDF is shown in Fig.\ref{Fig.5}a in a proper scaled form for $\sigma=0.25$. Data were obtained from all eigenfunctions with $-0.5<E<0.5$ with chains with distinct sizes and typically $10^3$ disorder realizations. According to RM theory, the wavefunction intensities shall follow the Porter-Thomas distribution \cite{porter}

\begin{equation}
	P(\phi^2) = \sqrt{\frac{N}{2\pi\phi^2}}e^{-N\phi^2/2} ,
	\label{Eq.4}
\end{equation}
which is shown as a solid curve in Fig.\ref{Fig.5}a and accurately fits the numerical data.

The wavefunction intensity distribution at the Anderson transition $\sigma=1$ is shown in Fig.\ref{Fig.5}b. The multifractal character of the critical wavefunctions leads to relevant deviations from the above Porter-Thomas form. Although a closed form of the critical wavefunction intensity distribution is not known, one can infer its asymptotic behaviors from the numerical data obtained from chains of distinct sizes. In the regime of very low intensities, the distribution also varies as $1/\sqrt{\phi^2}$. However, we find it displays a distinct finite-size dependence, being proportional to $N^{\beta}$, with $\beta = 0.64$. In the main frame of Fig.\ref{Fig.5}b we plot data in the proper scaling form  with data from distinct chain sizes collapsing into a single curve for low-intensities. Our results suggest the critical low-intensity wavefunction distribution to follow 
\begin{equation}
	P(\phi^2,N) = N^{\beta}\frac{g(\phi^2 N^{2\beta})}{\sqrt{\phi^2}} ,
	\label{Eq.6}
\end{equation}
with $g(x\rightarrow 0)$ approaching a constant and $\beta=0.64$. Notice that the above scaling function fails for large intensities. This reflects the need of distinct scaling exponents to characterize the underlying multifractality. In the inset of Fig.\ref{Fig.5}b we report our finite-size scaling result for the regime of large wavefunction-intensities. We evidence a slow convergence to the asymptotic exponential tail. Data collapse at the large-intensities distribution tail was  plotted in the scaling form


\begin{equation}
	P(\phi^2,N) = N^{-\eta}\frac{e^{-\phi^2N^{\gamma}}}{\sqrt{\phi^2}} .
	\label{Eq.7}
\end{equation}
with $\eta=1.2$ and $\gamma = 0.3$. The present analysis of the critical parent distribution adds to previous studies of scaling properties at the Anderson transition that focused on the wavefunction multifractal singularity spectrum and the inverse participation ratio distribution\cite{anderson3}. The power-law behavior at low wavefunction intensities is also consistent with previous reports signaling deviations from a log-normal distribution at the Anderson transition\cite{parent1,parent2}. To extract the multifractal singularity spectrum at the Anderson transition, we followed the prescription developed by Chhabra and Jensen\cite{jensen}. One firstly split the chain in $N_l=N/l$ segments of 
linear size $l$. For a given eigenfunction, we compute the probability $\mu_i$ to find the electron in the $i$th segment. After that, we introduce a set of generalized measures

\begin{figure}[t!]
	\centering
	\includegraphics[width = 0.49 \textwidth]{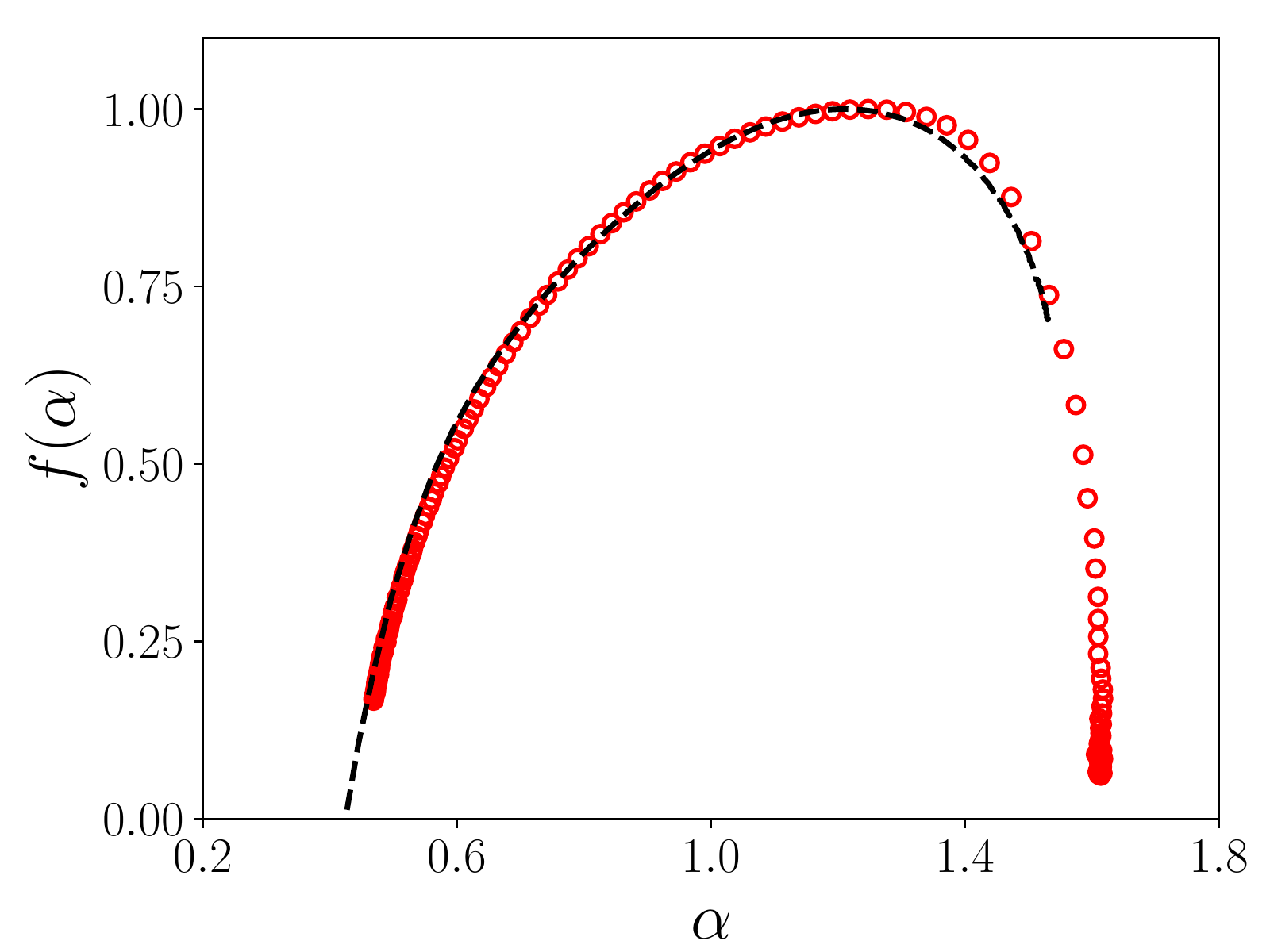}
	\caption{Multifractal singularity spectrum associated to the critical wavefunctions. Here, we used data from wavefunctions with energies in the interval $-1/2<E<1/2$. It closely satisfies the expected symmetry $f(2-\alpha)=f(\alpha)-\alpha+1$ shown as a dashed line. Data were obtained from chains with $N=1200$ sites splited in segments of size $l=2$ to reduce numerical uncertainties in the right branch of $f(\alpha )$.}
	\label{falpha}		
\end{figure}  

\begin{equation}
	\delta_i(q,N_l) = \frac{[\mu_i(N_l)]^q}{\sum_{i=1}^{N_l}[\mu_i(N_l)]^q}
\end{equation}
where $q$ is a parameter used to explore different sets of the multifractal measure $\delta_i(N_l)$. The Hausdorff fractal dimension of $\mu(q)$ can be expressed as
\begin{equation}
	f(q) = - \lim_{N_l\rightarrow\infty}\frac{1}{\ln{N_l}}\sum_{i=1}^{N_l}\delta_i(q,N_l)\ln{[\delta_i(q,N_l)]}.
\end{equation}
In addition, the average singularity strength with respect to $\delta(q)$ is given by 
\begin{equation}
	\alpha(q) = -\lim_{N_l\rightarrow\infty}\frac{1}{\ln{N_l}}\sum_{i=1}^{N_l}\delta_i(q,N_l)\ln{[\mu_i(q,N_l)]}.
\end{equation}
The function $f(\alpha)$, parameterized by $q$, gives the Hausdorff
fractal dimension of the set of points where $\mu$ scales as $N_l^{-\alpha}$.  Fig.\ref{falpha} shows our numerical result for the singularity spectrum averaged over all eigenfunctions with eigenenergies in the interval $-1/2<E<1/2$. It is consistent with previous reports of the multifractal singularity spectrum at the critical point of one-dimensional tight-binding models with power-law decaying hopping amplitudes in the regime of intermediate multifractality\cite{anderson3}. Further, it closely satisfies the well establish symmetry relation $f(2-\alpha)=f(\alpha)-\alpha+1$~\cite{mirlin2006}. We stress that the Hausdorff fractal dimension is strictly positive. Therefore, negative parts of $f(\alpha)$ related to rare events that typically do not occur in a single wavefunction are not reachable\cite{anderson3,romer1,romer2}.

Well within the phase of localized states, the wavefunction intensity distribution shows new trends, as illustrated in Fig.\ref{Fig.5}c on which we plotted $\phi^2 P(\phi^2)$ to clearly uncover four distinct regimes. For very small intensities $P(\phi^2)$ decays as $1/\sqrt{\phi^2}$. Although being size-independent, this behavior is similar to the initial decay of extended and critical states. Above a characteristic intensity that depends on the relative strength of diagonal and off-diagonal disorder, a faster than $1/\phi^2$ decay sets up. This regime is expected to appear due to the long-distance power-law decay of the wavefunction envelope. In this regime, the intensity distribution is also size-independent. The plateau depicted in Fig.\ref{Fig.5}c signals an $(N\phi^2)^{-1}$ decay of the intensity distribution which accounts for the exponential decay of the wavefunction profile at very-short distances $\phi^2(r)\propto e^{ -r/l_c}$ from which one can extract $P(\phi^2) =[N|d\phi^2/dr|]^{-1}$ \cite{extremecondensed2}. The deviation from the plateau for the largest intensities is associated to the emergence of  anomalously localized states.

\begin{figure}[p]
	\centering
	\includegraphics[width = 0.49 \textwidth]{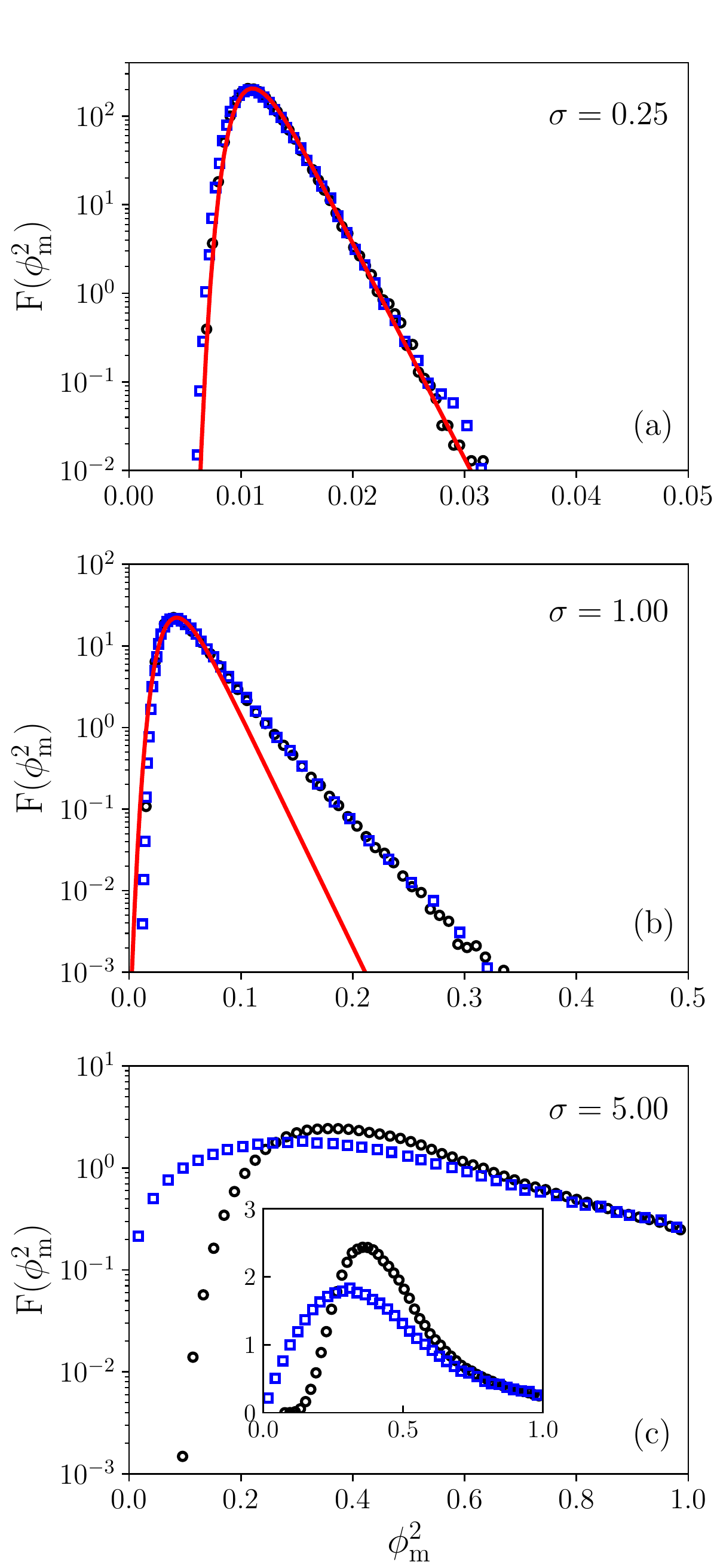}
	\caption{Wavefunction maximum intensity distributions for representative values of $\sigma$. (a) Well within the extended phase. The average maximum scales as $\ln N/N$ (see text) and the PDF has a Gumbel form (solid line). The measured distribution (circles) is well reproduced from the parent distribution assuming the IID hypothesis (squares). (b) At the Anderson transition. The average maximum scales as $N^{-\xi}$ with $\xi=0.413(2)$ (see Fig.\ref{Fig.4}). The measured distribution (circles) and the one derived from the IID hypothesis (squares) are slightly distinct and deviates from the Gumbel form (solid line). (c) Well within the localized phase. The distribution is size independent. The measured and IID-based distributions are similar in the strongly localized regime ($\phi_\mathrm{m}^2$ close to unit) but significantly deviate from each other as $\phi_\mathrm{m}^2$ decreases. }
	\label{Fig.6}		
\end{figure}

The wavefunction extreme value distributions of extended, critical and strongly localized states are reported in Fig.\ref{Fig.6}. Here we plot the numerically obtained distributions as well as the distributions derived from the parent distributions built in Fig.\ref{Fig.5} by assuming the IID variables hypothesis. For $\sigma=0.25$ (see Fig.\ref{Fig.6}a) both distributions are indistinguishable within the numerical accuracy. This feature supports the irrelevance of correlations in the structureless wavefunction profile according to the GOE statistics. The resulting distribution is fairly well fitted by the Gumbel distribution (Eq.\ref{Eq.3}) with parameters $a=559.95$ and $b=481.96$. By analysing data from distinct chain sizes, we found {that} the average maximum {scales with the same logarithmic correction displayed in the inset of Fig.\ref{Fig.4}, as expected for extended states.  The Gumbel distribution presented in Fig.\ref{Fig.6}a are the same for any $\sigma \leq 0.5$, where the system behavior is fully described by the GOE statistics}. 

{At criticality $\sigma = 1$, the wavefunction maximum distribution is not well fitted by the Gumbel curve. This feature is associated to the development of multifractal fluctuations \cite{long1,long2,long3}. To illustrate this behavior, we plot the measured distribution of maxima and the one assuming the IID hypothesis at the Anderson transition ($\sigma=1$ in Fig.\ref{Fig.6}b). As one can see, both distributions are numerically coincident but are not well fitted by the Gumbel curve. The peak of the extreme values distribution is at an intensity that scales as $N^{-\xi}$, the same scaling reported in Fig.\ref{Fig.4} for the average maximum intensity at criticality. }

\begin{figure}[t!]
	\centering
	\includegraphics[width = 0.49 \textwidth]{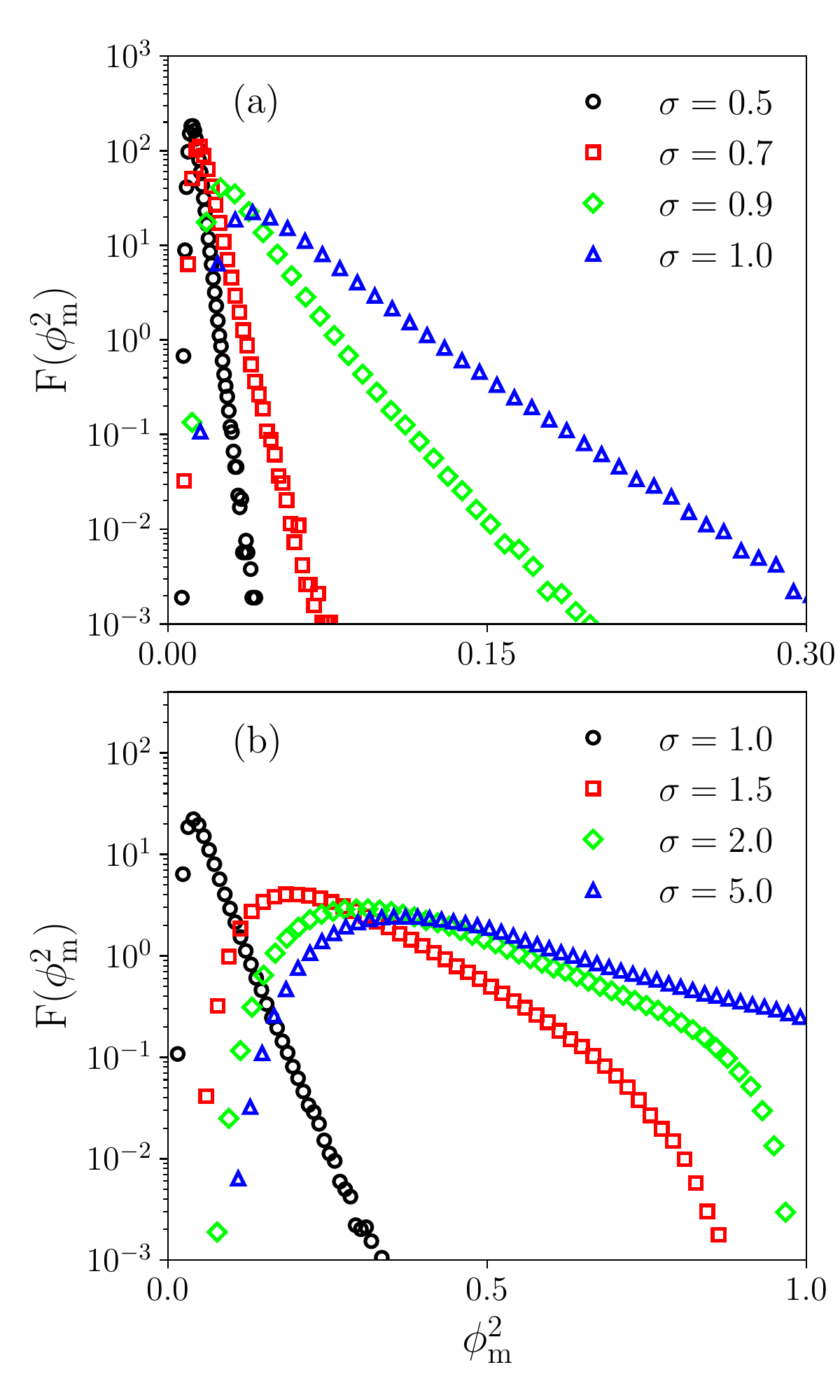}
	\caption{Wavefunction maximum distributions for distinct values of the hopping exponent $\sigma$. (a) In the extended regime, $\text{F}(\phi_{\mathrm m}^2)$ deviates from the Gumbel distribution when the wavefunction starts to display anomalous fluctuations ($1/2 < \sigma < 1$). (b) In the localized regime, the extreme distribution moves towards the asymptotic curve for high values of $\sigma$, as illustrated for $\sigma = 5.0$. }
	\label{Fig.7}
\end{figure}

{Right after the critical point, the IID hypothesis no longer holds for the system and then the data obtained from the derivative of Eq.\ref{Eq.2} starts to deviate from the measured one. In order to show this difference, we report the extreme distribution well within the localized phase ($\sigma=5$) in Fig.\ref{Fig.6}c. It is clear the different between those distributions, been similar only in the regime of very large maxima coming from the most localized states. The strong deviation from the IID prediction uncovers the underlying correlations emerging from the exponential short-distance wavefunction profile of typical localized states.   } 


{To finish, we illustrate in figure \ref{Fig.7} how the wavefunction maximum intensity distribution changes as the hopping exponent increases. In Fig.\ref{Fig.7}a, we plot the $\text{F}(\phi_{\mathrm m}^2)$ in the anomalous extended phase $1/2<\sigma<1$. As already mentioned, in the regime where the system is fully described by the GOE statistics ($0 < \sigma \leq 0.5$), the extreme distribution is well fitted by the Gumbel curve. When the extended wavefunctions start to develop anomalous fluctuations, the distribution  deviates significantly from Gumbel, showing a longer tail and a larger standard deviation. Fig.\ref{Fig.7}b shows the corresponding crossover from the critical to the well localized regime. As one enters the localized phase, the IID hypothesis fails. The distribution of wavefunction maxima becomes wider, exhibiting an intermediate pre-asymptotic regime. The exponential tail is fully suppressed in the regime of very short-ranged hopping amplitudes (large $\sigma$) with a finite probability density at $\phi_m^2 =1$ signaling the presence of extremely localized states.  }

\section{ Summary and Conclusion}

In summary, we considered the one-dimensional tight-binding Anderson model with random power-law decaying matrix elements  to uncover the extreme value distributions of extended, critical and exponentially localized wavefunctions. We recall that this model has been widely considered in the literature as a simple model that exhibits an Anderson
transition at a well-known critical point which has been used in several studies exploring
fundamental features related to the critical fluctuations\cite{long1,long2,long3,long4,long5,long6,long7,long8,long9,long10,long11,long12,long13,long14,long15,long16}.

We showed that structureless extended eigenstates satisfying the GOE ensemble exhibit a Gumbel distribution of intensity maxima that can be indeed extracted from the parent wavefunction Porter-Thomas distribution  assuming IID wavefunction values.  On the other hand, we unveiled that multifractal critical eigenstates have a nontrivial distribution of maxima whose tail deviates significantly from the Gumbel distribution due to the slow convergence to the exponential tail of the wavefunction distribution. 
The latter was shown to exhibit distinct scaling behaviors at low and large intensities. In particular, the average wavefunction intensity maximum behaves as an order-parameter measure. At criticality, it scales with the system size $\langle \phi^2_\mathrm{m}\rangle \propto N^{-\xi} $ with $\xi \simeq 0.413(2)$. 
Well within the localized phase, we showed that the wavefunction distribution exhibits a nontrivial sequence of regimes (pre-power-law $\rightarrow$ power-law $\rightarrow$ exponential $\rightarrow$ anomalous localized). This unconventional distributions was shown to impact on the extreme value distribution. The corresponding distribution of maximum eigenfunction intensities has contributions coming from the anomalously localized and exponential regimes. It substantially deviates from the distribution based on the IID hypothesis. Further, it depicts a finite probability density at $\phi^2=1$ coming from the anomalously localized states. We stress that, although the finite-size scaling exponents at the Anderson transition for the present model with long-range hopping amplitudes are non-universal, depending on the specific relation between short and long-range terms\cite{anderson3}, the main reported aspects related to the parent and maximum wavefunction distributions in the extended Gaussian, critical and exponentially localized regimes shall remain valid for general model systems presenting an Anderson transition. At criticality, the reported deviation from the standard Gumbel distribution of maxima shall also universally hold as it is directly related to the presence of multifractal wavefunction fluctuations. 

The detailed aspects of the new distributions reported in the present work were based on numerical results. We hope these can stimulate future developments based on the field-theoretical renormalization group and the nonlinear $\sigma$ model\cite{long1,long2,long3} with the potential to derive analytical forms for these distributions at least in some relevant asymptotic regimes. The here disclosed features of the wavefunction extreme value distributions open a new approach to investigate the Anderson transition. Questions related to the influence of mobility edges, system's dimensionality, interactions, and class of random matrix ensemble deserve future investigations. Considering that extreme wavefunctions play a key role in disordered systems, these studies shall shed new light on the general wave transport phenomenology.

\section{ Acknowledgement}	
This work was supported by CAPES (Coordenação de Aperfeiçoamento de Pessoal de Nível Superior), CNPq (Conselho Nacional de Desenvolvimento Científico e Tecnológico) and FAPEAL (Fundação de Apoio à Pesquisa do Estado de Alagoas).


\end{document}